\documentclass{article}
\usepackage[dvips]{graphics}

\@ptsize
\or 
\or 
\fi \advance\textheight by \topskip

\catcode`\@=12

\begin{document}
\title{\bf{Study of the temperature dependence of single-particle
and pair coherent condensate
densities for a Bose liquid with a "depleted" single-particle
Bose-Einstein condensate (BEC) at $T \neq 0$}}
\author{E.Pashitskii
\thanks{E-mail: pashit@iop.kiev.ua}\qquad
S.Vilchynskyy \thanks{E-mail: sivil@mail.univ.kiev.ua}\\
\phantom{dfghdfhrrrj}\\
\it Institute of Physics, NAS of Ukraine, Kiev 03022, Ukraine;\\
 \it Taras Shevchenko National Kiev University, Kiev 03022, Ukraine}
\date{}
\maketitle
\begin{abstract}
We study the temperature dependence of single-particle and pair
coherent condensate densities for a Bose liquid with a
"depleted" single-particle Bose-Einstein condensate (BEC) at
$T\neq0$. Our investigations were based on the field-theoretical
Green's functions. On the basis of the use of empirical data about
the speeds of the first and second sounds  is described superfluid
state at $T\neq 0.$ It is studied the structure of superfluid
state taking into account appearance with the different from zero
temperatures of normal component and taking into account the
branch of the second sound, whose speed approaches zero at $T\to
T_\lambda$.  The bare interaction between bosons was chosen in the
form of a repulsive Azis potential, the Fourier component of which
was an oscillatory and sign-varying function of the momentum. The
obtained temperature dependence of single-particle and pair
coherent condensate and total densities has a good agreement with
experiment data.
\end{abstract}
\newpage
\section{Introduction}
    Despite the big progress achieved in the theory of superfluidity
since the pioneer works by Landau \cite{LDL}, Bogolyubov
\cite{BNN}, Feynman \cite{FR} and others \cite{N-P}-\cite{5.30},
the task of constructing a microscopic theory of a superfluid (SF)
state of a $^4$He Bose liquid cannot be considered complete.
Analysis of experimental and theoretical activities concerning with
unique superfluidity phenomena investigation has one conclude that
microscopic theory of superfluidity is still incomplete. There are
some unsolved contradictions between theory and experiment as
concerned with superfluid liquid hydrodynamics as with structure
of quasiparticle  spectrum. These contradictions reduce to the
question of quantum-mechanical structure of superfluid ${}^4$He
component below the $\lambda$-point. This question is crucial in
creation of consistent microscopic superfluid theory of Bose
liquid both for the case of zero temperature and for the case of
different from zero temperatures.

    In works \cite{PRL}, \cite{JLTP} one introduces
at $T=0$ the microscopic model of the SF state  of a Bose liquid
with a single-particle Bose-Einstein condensate (BEC) suppressed
because of interaction, based upon a renormalized field
perturbation theory with combined variables
\cite{53.10}--\cite{53.32}, allows one to obtain a self-consistent
``short'' system of nonlinear integral equations for the
self-energy parts $\tilde\Sigma_{ij}(p,\epsilon)$, by means of
truncating the infinite series in the small density of the BEC
($n_0/n\ll 1$). By the same token, one can work out a
self-consistent microscopic theory of a superfluid Bose liquid and
perform an ab initio calculation of the spectrum of elementary
excitations $E(p)$, starting from realistic models of pair
interaction potential $U(r).$

    In this work we will study of the superfluid state structure of
Bose-liquid ${}^4$He under nonzero temperature $(0<T<T_\lambda).$
The same task was consider in work \cite{germ} on the basic of
Chester \cite{chester} IBG wave function $\Psi_{IBG}$ by Jastrow
factors $F=\Pi f_{ij}$.
 We  use the microscopic model proposed in \cite{53.17} for
  the superfluid  Bose liquid with a "depleted"  BEC.
  The small parameter
   of these model is the ratio of the density of the BEC to the total
    density Bose liquid $\rho_0/\rho_s \ll 1$ in contrast
    to the Bogolyubov
 theory \cite{BNN} for a almost ideal Bose gas, where the small
 parameter is the ratio of the number of overcondensate
 excitations to the number of particles in the intense BEC
$(n-n_0)/n_0\ll 1.$

Such approach is based on the analysis of the numerous precise
experimental data on neutron inelastic scattering \cite{5.30},
\cite{15}--\cite{17} and to results in quantum evaporation of
$^4$He atoms \cite{19}, according to which the maximal density
$\rho_0$ of the single-particle BEC in the $^4$He Bose liquid even
at very low temperatures $T\ll T_\lambda$ does not exceed $10 \%$
of the total density $\rho$ of liquid $^4$He, whereas the density
of the SF component $\rho_s \to \rho $ at $T\to 0$ \cite{53.23}.
Such a low density of the BEC is implied by strong interaction
between $^4$He atoms and is an indication of the fact that the
quantum structure of the part of the SF condensate in He~II
carrying the ``excess'' density $(\rho_s-\rho_0)\gg \rho_0$ calls
for a more thorough investigation. So, in this case the density of
the superfluid component $\rho_s$  is determined by the quantity
of the renormalized anomalous self-energy parts $\tilde
\Sigma{}_{12}(0,0)$, which is a superposition of the "depleted"
single-particle BEC and the intense "Cooper" pair coherent
condensate (PCC), with coincident phases (sings) of the
corresponding order parameters. Such a PCC emerges due to an
effective attraction between bosons in some regions of momentum
space, which results from an oscillating sign-changing momentum
dependence of the Fourier component $V(p)$ of the interaction
potentials  $U(r)$ with the inflection points in the radial
dependence.

    The system of Dyson-Belyaev equations \cite{6}, which
describe the superfluid (SF) state of a Bose liquid with strong
interaction between bosons and a weak single-particle
Bose-Einstein condensate (BEC) and which allows one to express the
normal $\tilde G_{11}$ and anomalous $\tilde G_{12}$ renormalized
single-particle boson Green functions in terms of the respective
self-energy parts $\tilde \Sigma_{11}$ and $\tilde \Sigma_{12}$,
has the next analytic form of \cite{53.17}:

\begin{equation}
\tilde \Sigma_{11}(\mathbf{p}, \epsilon)= n_0\Lambda (\mathbf{p},
\epsilon)
 \tilde V(\mathbf{p}, \epsilon)+
 n_1 V(0)+\tilde \Psi _{11}(\mathbf{p}, \epsilon)\;;
\label{a2}
\end{equation}
\begin{equation}
\tilde \Sigma_{12}(\mathbf{p}, \epsilon) = n_0\Lambda (\mathbf{p},
\epsilon) \tilde V(\mathbf{p}, \epsilon)+ \tilde \Psi_{12}
(\mathbf{p}, \epsilon)\;, \label{a3}
\end{equation}
where
\begin{equation}
 \tilde{\Psi}_{ij}(\vec
p,\epsilon)=i\int\frac{d^3 \vec k}{(2\pi)^3}
 \int\frac{d\omega}{2\pi}G_{ij}(\vec k)\tilde V(\vec p-\vec k,\epsilon-\omega)
 \Gamma(\vec p,\epsilon,\vec k,\omega)
 \label{a4}
 \end{equation}
\begin{equation}
\tilde V(\vec p,\epsilon)=V(\vec p)\left[1- V(\vec p)\Pi(\vec
p,\epsilon)\right]^{-1}.
 \label{a5}
\end{equation}
where $V(p)$ is the  Fourier component of the bare potential of
the pair interaction of the bosons; $\tilde V (\vec p,\epsilon)$
is the renormalized (screened) pair interaction between bosons
because of  many-particle collective effects; $\Pi(\vec
p,\epsilon)$ is the boson's polarization operator which describe
many-particle correlation  effects
\begin{equation}
\begin{array}{l}
\displaystyle \Pi(\mathbf{p}, \epsilon)= i\int \frac{d^3
\mathbf{k}}{(2\pi)^3}\int \frac{d\omega}{2\pi}\,
\Gamma(\mathbf{p}, \epsilon,\mathbf{k},\omega)
\left\{G_{11}(\mathbf{k}, \omega)\right.
\\[12pt]
{}\quad\times\left. G_{11}(\mathbf{k}+\mathbf{p}, \epsilon+\omega)
+G_{12}(\mathbf{k}, \omega) G_{12}(\mathbf{k}+\mathbf{p},
\epsilon+\omega)\right\}\;;
\end{array}
\label{a6}
\end{equation}
where $\Gamma(\vec p,\epsilon;\vec k,\omega)$ is the vertex part
(three-pole) describing many-particles correlations of the local
field's type; $\Lambda(\vec p, \epsilon)=\Gamma(\vec p,
\epsilon,0,0)=\Gamma(0,0,\vec p, \epsilon)$ is the vertex part
with zero values of input momentum and energy; and $n_0$ is the
number particles in BEC,  $n_1=n-n_0$ is the number of
overcondensate particles \mbox{$(n_1 \gg n_0),$} which is
determined by the condition of the total particle number
conservation:
\begin{equation}
n=n_0+n_1=n_0+ i\int\frac{d^3 \vec k}{(2\pi)^3}\int
\frac{d\omega}{2\pi}
 G_{11}(\vec k, \omega).
\label{a7}
\end{equation}
Take into account the residues at the poles of single-particle
Green functions $\tilde G_{ij}(\mathbf{p}, \epsilon)$, neglecting
the contributions of eventual poles of the functions $\Gamma
(\mathbf{p},\epsilon, \mathbf{k}, \omega )$ and $\tilde
V(\mathbf{p}, \epsilon)$, which do not coincide with the poles of
$\tilde G_{ij}(\mathbf{p}, \epsilon)$, the functions $\tilde
\Psi_{ij}(\mathbf{p},\epsilon)$ on the ``mass shell''
$\epsilon=E(p)$ assume the following form (at $T=0$):
\begin{equation}
\begin{array}{l}
\displaystyle
 \tilde{\Psi}_{11}(\mathbf{p}, E(p))=
\frac{1}{2}\int \frac{d^3 \mathbf{k}}{(2\pi)^3} \Gamma
(\mathbf{p}, E(p); \mathbf{k}, E(k))
 \\[12pt]
\displaystyle {} \quad \times \tilde{V}(\mathbf{p}- \mathbf{k},
E(p)- E(k)) \left[\frac{A(\mathbf{k}, E(k))} {E(k)}-1\right]\;;
\end{array}
\label{a8}
\end{equation}
\begin{equation}
\begin{array}{l}
\displaystyle \tilde{\Psi}_{12}(\mathbf{p},  E(p))=
 -\frac{1}{2}\int \frac{d^3  \mathbf{k}}
{(2\pi)^3}\\[12pt]
\displaystyle {} \quad \times \tilde{V}(\mathbf{p}- \mathbf{k},
E(p)- E(k)) \Gamma (\mathbf{p}, E(p);
\mathbf{k}, E(k))\\[12pt]
\displaystyle {} \quad \times\frac{n_0\Lambda (\mathbf{k}, E(k))
\tilde{V}(\mathbf{k}, E(k))+
\tilde{\Psi}_{12}(\mathbf{k},E(k))}{E(k)}\;,
\end{array}
\label{a9}
\end{equation}
where
\begin{equation}
A(\vec p)=n_0\Lambda (\vec p, E(\vec p))\tilde{V}(\vec p, E(\vec
p))+
 n_1V(0) +\tilde{\Psi}_{11}(\vec p) + \frac{\vec p\,^2}{2m}-\mu,
 \label{a10}
\end{equation}
here $\mu$ is the chemical potential of the quasiparticles, which
satisfies the Hugengoltz-Pines relation \cite{26}
\begin{equation}
\mu=\tilde\Sigma_{11}(0,0)-\tilde\Sigma_{12}(0,0)\;. \label{a1a}
\end{equation}
 The quasiparticle spectrum corresponding to the poles of Green's
function  is determined in general, with allowance for relations
(\ref{a2},\ref{a3}),  by the next expression
\begin{equation}
E(\vec p)=\sqrt{A^2(\vec p)-\left[
 n_0\Lambda(\vec p, E(\vec p)) \tilde{V}(\vec p, E(\vec p))+
 \tilde{\Psi}_{12}(\vec p)\right]^2}.
 \label{a11}
\end{equation}

\section{ The structure of the superfluid Bose liquid state at
$T\neq 0$ }
    The main purpose of our research is the investigation
of superfluid helium condensate structure by means of finding
temperature dependence  of single-particle and pair coherent
condensate densities under assumption that superfluid component is
a superposition of a weak single-particle BEC and an intensive
PCC.

    Let's consider the superfluid bose liquid state at $T\neq0$ when
there are superfluid component density $\rho_s(T)$ and normal
component density $\rho_n(T)$ simultaneously. As it was shown in
\cite{53.31,53.32} the expression for the renormalized Green's
function $\tilde G{}_{ij}(p)$ is constructed with help of combined
variables:
\begin{equation}
\tilde \Psi(x)= \tilde \Psi_L (x)+\tilde \Psi_{sh} (x). \label{b1}
\end{equation}
In the long-wavelength region $|\vec k|<k_0$ (where $k$ is some
characteristic momentum) this variables are just the hydrodynamics
variables $\tilde \Psi{}_L(x)$ in the spirit of Landau quantum
hydrodynamics \cite{LDL}, and  in the short-wavelength region
$|\vec k|>k_0$  they coincide with the usual field operators
$\tilde \Psi{}_{sh}(x)$:
\begin{equation}
\begin{array}{c}
 \tilde \Psi_\mathrm{L} (x)= \sqrt{\left\langle \tilde
n_\mathrm{L}\right\rangle}\left[ 1+\frac{\tilde
n_\mathrm{L}-\left\langle \tilde n_\mathrm{L}\right\rangle}
{2\left\langle \tilde n_\mathrm{L}\right\rangle} +i\tilde
\phi_\mathrm{L} \right]; \quad \tilde \Psi_\mathrm{sh}=
\psi_\mathrm{sh}e^{-i\tilde\phi_\mathrm{L}}\;;\\[12pt]
\psi_\mathrm{sh}=\psi-\psi_\mathrm{L};\quad
\psi_\mathrm{L}(\mathbf{r})=\frac{1}{\sqrt{V}}
\sum_{|\mathbf{k}|<k_0} a_{\mathbf{k}} e^{i\mathbf{k} \mathbf{r}}=
\sqrt{\left\langle \tilde n_\mathrm{L}\right\rangle} e^{i\tilde
\phi_\mathrm{L}}\;.
\end{array}
\label{b2}
\end{equation}

Such an approach means that the separation of the Bose system into
a macroscopic coherent condensate and a gas of supracondensate
excitations is made not on the statistical level, like in the case
of a weakly nonideal Bose gas \cite{BNN}, but on the level of ab
initio field operators, which are used to construct a microscopic
theory of the Bose liquid.

    In the region of small $p\neq0$ at $T\rightarrow0$ Green's
functions have the form
\begin{eqnarray}\label{b3}
\tilde
G_{11}(p)=-n_sg_{\varphi\varphi}(p)-ig_{\varphi\pi}(p)-\phantom{Tan}\nonumber
\\
-\frac1{4\rho_s}g_{\pi\pi}(p)-\frac{n_s}2\Phi_{\varphi\varphi}(p)\ldots
\end{eqnarray}
\begin{eqnarray}\label{b4}
\tilde G_{12}(p)=n_sg_{\varphi\varphi}(p)
-\frac1{4\rho_s}g_{\pi\pi}(p)-\frac{n_s}2\Phi_{\varphi\varphi}(p)\ldots
\end{eqnarray}
where
\begin{eqnarray}\label{b5}
\Phi_{\varphi\varphi}(p)=\int_{|q|<q_0}\frac{d^4q}{2\pi^4}g_{\varphi\varphi}(q)
g_{\varphi\varphi}(p-q),\nonumber\\
p=(\vec k,\epsilon), q=(\vec q,\omega)\phantom{gdfbdskjfgbkjd\,r}
\end{eqnarray}
where $g_{\mu\nu}(p)$ are the "hydrodynamics" Green's functions,
which are associated with the long-wavelength fluctuations of the
phase and density of the condensate ( $\mu\nu=\varphi\pi$). The
expressions for
$g_{\varphi\varphi}(p),g_{\varphi\pi}(p),g_{\pi\pi}(p)$ was
calculated in \cite{53.11} for $T>0$. It contains sums of two pole
terms, corresponding to first and second sound with velocities
$c_1$ and $c_2$ in the Bose liquid with the normal and superfluid
components:
\begin{eqnarray}\label{b6}
g_{\mu\nu}(k,\epsilon)=\frac{(a_{\mu\nu}-d_{\mu\nu}\rho_n/\rho)}
{\epsilon^2-c_1^2k^2}+\frac{b_{\mu\nu}\,\rho_n/\rho}{\epsilon^2-c_2^2k^2}\nonumber\\
\mu,\nu=\varphi,\pi \phantom{fdgfhrsthyjuytjiuytdjmnyt}
\end{eqnarray}
where $\rho=\rho_n+\rho_s$ is the total density of the liquid, and
the coefficients $a_{\mu\nu}, d_{\mu\nu}$ and $b_{\mu\nu}$ are
independent of $T$ at low temperatures. This result establishes a
unique correspondence between the microscopic field theory of
superfluidity \cite{6,7} and the macroscopic two-fluid
hydrodynamics \cite{20,53.23}.

It follows from Eqs. (\ref{b3}), (\ref{b4}) and (\ref{b6}) that
the pole parts of the renormalized Green's functions $\tilde
G{}_{ij}$ can be represented in the form
\begin{eqnarray}\label{b7}
\tilde G {}_{ij}(\vec
k,\epsilon)=\frac{(A_{ij}-D_{ij}\rho_n/\rho)}
{\epsilon^2-c_1^2k^2}+\frac{B_{ij}\,\rho_n/\rho}{\epsilon^2-c_2^2k^2}
\qquad i,j=1,2
\end{eqnarray}

We henceforth assume that expression (\ref{b7}) is valid in the
entire temperature interval $T<T_\lambda$. It follows from Eqs.
(\ref{b7}) that for $T\rightarrow0$, where $\rho_n\rightarrow0$,
the leading contribution to the integral over energy $\epsilon$ in
(\ref{a4}) comes from the first-sound pole $\epsilon = c_1 k$ of
the Green's functions. However, at higher temperatures $T>1K$,
where $\rho_n \sim \rho_s$,
 the main part,
because of the strong inequality $c_1 \gg c_2$, is played by the
low-energy pole $\epsilon = c_2k$ which corresponds to second
sound.

At finite temperatures $(T\neq 0)$, after taking into account the
contributions of the first and second poles of the Green's
functions (\ref{b7}), we obtain for the self-energy parts in
$\Gamma=\Lambda=1.5$ approximation
\begin{eqnarray}\label{b8}
\tilde \Sigma{}_{ij}(\vec k,T)=-\frac12\int \frac{d^3\vec
q}{(2\pi)^3}\tilde V(\vec k -\vec
q)\left\{\left[A_{ij}-D_{ij}\frac{\rho_n(T)}\rho\right]\frac1{c_1q}
\right. \nonumber\\
\left.\times
\coth\left(\frac{c_1q}{2T}\right)+B_{ij}\frac{\rho_n(T)}\rho\frac1{c_2q}
\coth\left(\frac{c_1q}{2T}\right)\right\}\phantom{dfg}
\end{eqnarray}

It should be emphasized that the long-wavelength approximation for
the Green's functions (\ref{b7}) in this case are valid because of
the divergence of the temperature factor
     $\coth(c_2q/2T)$  at  $q\rightarrow 0$  and the rather rapid decay
of the interaction kernel    ( $q\rightarrow \infty$). Moreover,
the system of equations (\ref{b8}) does not need to be matched
with the expression for the renormalized quasi-particle spectrum
$E(k)$, as it is ordinarily done in microscopic field theory for
$T\rightarrow 0$. It's true because the substitution of the
empirical spectra of the first and second sound (with the
experimental values of the velocities $c_1$ and $c_2$) into the
expressions for the Green's functions $\tilde G{}_{ij}(p)$
corresponds to automatically taking all the necessary
renormalizations into account.

 Using (\ref{b8}), one can determine the superfluid order
parameter for $T\neq0$:
\begin{eqnarray}\label{b9}
\tilde \Sigma{}_{12}(0,T) =
\Psi_0(T)+\Psi_s(T)\frac{\rho_s(T)}\rho,
\end{eqnarray}
where
\begin{eqnarray}\label{b10}
\Psi_0(T)=-\frac12\int \frac{d^3\vec q}{(2\pi)^3}\tilde V(q)
\left[\frac{A_{12}-D_{12}}{c_1q}\coth\left(\frac{c_1q}{2T}\right)
+\frac{B_{12}}{c_2q}\coth\left(\frac{c_2q}{2T}\right)\right],
\end{eqnarray}
\begin{eqnarray}\label{b11}
\Psi_s(T)=-\frac12\int \frac{d^3\vec q}{(2\pi)^3}\tilde V(q)
\left[\frac{D_{12}}{c_1q}\coth\left(\frac{c_1q}{2T}\right)
-\frac{B_{12}}{c_2q}\coth\left(\frac{c_2q}{2T}\right)\right],
\end{eqnarray}

 Since under $T=0$ the density  of the superfluid component
 $\rho_s$ coincides with the total mass density of the Bose liquid
 $\rho=mn$ and is proportionate to $\Sigma_{12}(0,0)$ (it  plays a
 part of
 superfluid order parameter) we can write the following expression
\begin{eqnarray}\label{b12}
\rho_s=\rho_0+\tilde \rho{}_s=\beta m
\frac{\Sigma_{12}(0)}{\Lambda(0)\tilde V(0)}=\beta m [n_0
(1-\gamma)+\theta],
\end{eqnarray}
where
\begin{eqnarray}\label{b13}
\gamma\equiv\frac1{(2\pi)^2\Lambda(0)\tilde
V(0)}\int\limits_0^\infty
\frac{k^2\,dk}{E(k)}\left(\Lambda(k)\tilde V(k)\right)^2,
\end{eqnarray}
\begin{eqnarray}\label{b14}
\theta=-\frac1{(2\pi)^2\Lambda(0)\tilde V(0)}\int\limits_0^\infty
\frac{k^2\,dk}{E(k)}\Lambda(k)\tilde V(k)\Psi(k),
\end{eqnarray}
 where  $\beta$ is the constant to be identified. Starting from the
 definition of the single-particle BEC  density $\rho_0$ as
 $\rho_0 =mn_0$, we obtain $\beta=(1-\gamma)^{-1}$. Thus the
 "Cooper"   PCC density takes form:
\begin{eqnarray}\label{b15}
\tilde \rho{}_s=mn_1=\frac{m\theta}{1-\gamma},
\end{eqnarray}
 where $\rho_0 =mn_0$ is the density of the single-particle BEC
 and $\tilde \rho{}_s$ is the density of "Cooper"  PCC.

 On the other hand, the temperature dependence of BEC density
 $\rho_0(T)=mn_0(T)$ is due to entire particle number
 conservation condition (\ref{a6})
 will be determined the following expression
\begin{eqnarray}\label{b17}
\frac{\rho_0(T)}\rho=1-\frac12\int \frac{d^3q}{(2\pi)^3}
\left\{\left[A_{11}-D_{11}\frac{\rho_n(T)}\rho\right]\frac1{c_1q}
\right. \phantom{df}\nonumber\\
\left.\times
\coth\left(\frac{c_1q}{2T}\right)+B_{11}\frac{\rho_n(T)}\rho\frac1{c_2q}
\coth\left(\frac{c_2q}{2T}\right)\right\}.
\end{eqnarray}

 The density of pair coherent condensate is
\begin{eqnarray}\label{b18}
\frac{\rho_s(T)}\rho=\frac{\Psi_0(T)}{\tilde
V(0)n}\left[1-\frac{\Psi_s(T)}{\tilde V(0)n}\right]^{-1}.
\end{eqnarray}

 Thus the "Cooper" PCC density at $T=0$ in our model can be found in two
 independent ways. We can find it directly with help of expression (\ref{b15}) or
we can  subtract (\ref{b17}) from (\ref{b18}) and then find result
at $T=0.$ Thus we can verify the self-consistency of the
 proposed model via comparison the results of independent calculations
 of the PCC density of superfluid component for the zero temperature.

\section{The iterative scheme of calculation and  results}
 In order to calculate the the
temperature dependence of single-particle and pair coherent
condensate densities within the model of a Bose liquid with a
suppressed BEC considered, at first, we have to find  the  Fourier
component of the bare potential of the pair interaction of the
bosons and then, to find  the renormalized (screened) pair
interaction between bosons, which is determined (\ref{a5}), so we
must to calculate the boson polarization operator (\ref{a6}).

     The pair interaction between bosons was chosen in the
form of a regularized repulsive potential in the the Aziz
potential \cite{3,4}:
\begin{equation}
U_\mathrm{A}(r) = \left\{
\begin{array}{ll}
A\exp(-\alpha r - \beta r^2) -
\exp\left[-(r_0/r-1)^2\right]\sum_{k=0}^2
c_{2k+6}r^{-2k-6}\;,& r < r_0\,\\
A\exp(-\alpha r - \beta r^2) - \sum_{k=0}^2 c_{2k+6}r^{-2k-6}\;,&
r \ge r_0 \;
\end{array}
\right. \label{36}
\end{equation}
where $A=1.8443101\times10^5$~K, $\alpha=10.43329537$~\AA$^{-1}$,
$\beta=2.27965105$~\AA$^{-2}$, $c_6 = 1.36745214\mbox{
K}\times\mbox{\AA}^6$, $c_8 = 0.42123807\mbox{
K}\times\mbox{\AA}^8$, $c_{10} = 0.17473318\mbox{
K}\times\mbox{\AA}^{10}$. Such potential  remain finite at $r=0$
due to the nonanalytic exponential dependence on $r$, which
suppresses any power divergence at $r\to0.$ It has the Fourier
component $V(p),$ which is an oscillatory and sign-varying
function of the momentum $p$ as a result of the "excluded volume"
effect and the quantum diffraction of the particles on one
another.

    Then, in order to calculate the polarization operator, using
Eqs.~(\ref{a8}) and (\ref{a9}), a numerical calculation in the
first approximation of the functions $\Phi_1(p)\equiv\tilde
\Psi_{11}(\mathbf{p},E_0(\mathbf{p}))$ and $\Psi_1(p)\equiv
\tilde\Psi_{12}(\mathbf{p},E_0(\mathbf{p}))$, was conducted. For
the zeroth approximation,  the ``screened'' potential (\ref{a5})
were taken:
\begin{equation}
\tilde V_0(p)=\frac{V_0j_1(pa)}{pa-V_0\Pi_0 j_1(pa)} \label{51}
\end{equation}
at some constant negative value of $\Pi_0$. Then, using the
functions $\Phi_1(p)$ and $\Psi_1(p)$ obtained, the first
approximation for the polarization operator $\Pi_1(p)$ was
calculated, using Eqs.~(\ref{b3})--(\ref{b5}), (\ref{b7}) at
$\Gamma=1$.  The limiting value $\Pi_1(0)$ was compared with the
exact thermodynamic value of the polarization operator of the
$^4$He Bose liquid at $p=0$ and $\omega=0$, which determines the
compressibility of the Bose system \cite{53.31}: $ \displaystyle
\Pi(0,0)=-\frac{n}{mc_1^2}\;.$

The absolute value $\vert\Pi(0,0) \vert$ turned out to be almost
1.5 times greater than the calculated value $\vert\Pi_1(0)\vert$.
This provides for an estimate of the vertex $\Gamma$ at $p=0$ in
the first approximation as $\Gamma_1\equiv\Lambda_1\simeq 1.5$.
The second approximation $\Phi_2(p)$ and $\Psi_2(p)$ was obtained
from Eqs.~(\ref{a8}), (\ref{a9}) with the constant value
$\Gamma_1\equiv\Lambda_1$ and the first approximation for the
renormalized pseudopotential, Eq.~(\ref{a5}):
\begin{equation}
\tilde V_1(p)=\frac{V_0j_1(pa)}{pa-V_0\Pi_1(p)j_1(pa)}\;.
\label{52}
\end{equation}
Such an iterative procedure was repeated four to six times and
used to improve precision in the calculation of the polarization
operator. At each stage, equations (\ref{a10}) and (\ref{a11})
were used to reproduce the quasiparticle spectrum $E(p)$, and the
rate of convergence of the iterations was watched, as well as the
 degree of proximity of $E(p)$
to the empirical spectrum $E_\mathrm{exp}(p)$. The fitting
parameter in these calculations was the amplitude $V_0$ of the
seed potential (\ref{36}) at the value of $a=2.44$~\AA, which is
equal to twice the quantum radius of the $^4$He atom. The BEC
density, in accordance with the experimental data
\cite{15}--\cite{19}, was fixed at $n_0=9\% n=1.95\cdot
10^{21}\,\mathrm{cm}^{-3}$.

    For the numerical computation of temperature dependencies of
single-particle and pair coherent condensate density we have to
take into account that the first (hydrodynamics) sound velocity is
practically independent of $T$ and in the given approximation can
be determined as $c_1=\left[\tilde V(0)n/m^{*}\right]^{1/2}$( here
$m^*$ is the effective mass of quasiparticles), whereas the
velocity of second
 sound $c_2$ is substantially $T$-dependent, varying from $c_2(0)
 =c_1/\sqrt{3}$ at $T=0$ to a value $c_2(T)\simeq 20$m/s in the
 region $T>1K$ \cite{20}. At  $T\rightarrow T_\lambda$ the velocity
 $c_2\rightarrow 0$. Thus as the $\lambda$ point is approached
 the main part,
  owing to the strong inequality $c_1\gg c_2$, begins
 to be played by the last terms in the integrands in (\ref{b10}), (\ref{b17}),
 which are proportional to $B_{12}$ and $B_{11}$ and contain the
 temperature factor
\begin{eqnarray}\label{b19}
f(q,T)=\frac1{c_2(T)q}\coth\left(\frac{c_2(T)q}{2k_BT}\right)=
\frac{2T}{c_2^2(T)q^2}\, ,\qquad c_2q<T
\end{eqnarray}
 which diverges quadratically as $q\rightarrow0$. At the same time the width
 of the singular peak increases rapidly with increasing $T$ and
 decreasing $c_2$.
(the momentum dependence of temperature factor, obtained for
different temperatures, is depicted in fig.~1).
 As a result, with increasing $T$ there is an increase of the
 contribution to the integral (\ref{b10}) from the repulsive part of
 the potential $\tilde V(q)>0$ in the long-wavelength region $q<\pi/a$
 and a decrease of the function $\Psi_0(T)$.
 The function $\Psi_0(T)$  plays a part
 of the superfluid order parameter and  is positive at low
 $T<c_2\,q$ owing to the strong attraction $\tilde V(q)<0$
 in the region $\pi/a<q<2\pi/a$.
 At a certain critical temperature $T=T_c$ the function
$\Psi_0(T)$ goes to zero and then becomes negative (for $T>T_c$),
that corresponds to destruction of the superfluid state $(\rho_s
=0)$, i.e. $T_c$ coincides with the $\lambda$ point.

In a similar way with increasing $T$, there is increase of the
negative  contribution to the  integral (\ref{b17}) and decrease
of  $\rho_0(T)$  until at a certain point $T=T_0$ the density of
the BEC vanishes  and formally becomes negative for $T>T_0$.

    The results of the numerical calculations of the temperature
dependence of the densities of single-particle and paired parts of
superfluid density in the range of temperatures from zero to the
point of lambda-transition, are depicted on fig.2.

\section{Conclusions}

    Thus, on the basis of the use of empirical data about the speeds
of the first and second sounds in this work is described
superfluid state at $T\neq 0.$ It is studied the structure of
superfluid state taking into account appearance with the different
from zero temperatures of normal component and taking into account
the branch of the second sound, whose speed approaches zero at
$T\to T_\lambda$. We obtained the analytic expression (\ref{b17},
\ref{b18}) for computation $\rho_0$ and $\rho_s$ densities. and
constructed and realized the numerical scheme for computation of
the temperature dependence of single-particle and pair coherent
condensate densities.  The temperature dependence of the densities
of single-particle condensate and total density of superfluid
component are obtained on the basis of the microscopic model of
superfluid state of Bose-liquid with the depressed single-particle
BEC in the range from $T=0$ to temperatures  close to the
environment of point of $\lambda$-transition.

    Note also that the self-consistency of this model is corroborated
by the fact that the theoretical value of total particle density
calculated from Eq.~(\ref{a7}), $n_\mathrm{th}=2.1\cdot
10^{22}\mbox{ cm}^{-3}$, is quite close to the experimental value
of the density of particles in liquid $^4$He, $n=2.17\cdot
10^{22}\mbox{ cm}^{-3}$ (at $n_0=9\%\, n$). On the other hand, the
density $n_1$ of supracondensate particles, calculated from
Eqs.~(\ref{b15}) at $T=0$ at the values of the parameters
indicated, is about $0.93\,n$, which is also in good accordance
with experiment, taking into account that the BEC density is
determined up to $\pm 0.01\,n$.

    It goes without saying, the given results, which correspond to the
approximation of self-consistent field, cannot be used directly
near the $\lambda$-point, where thermodynamic fluctuations play
the key role. But the dependences, depicted in fig.2,
qualitatively correctly describe the temperature dependence of the
density of superfluid component.

\newpage

\begin{figure}
\begin{center}
\includegraphics{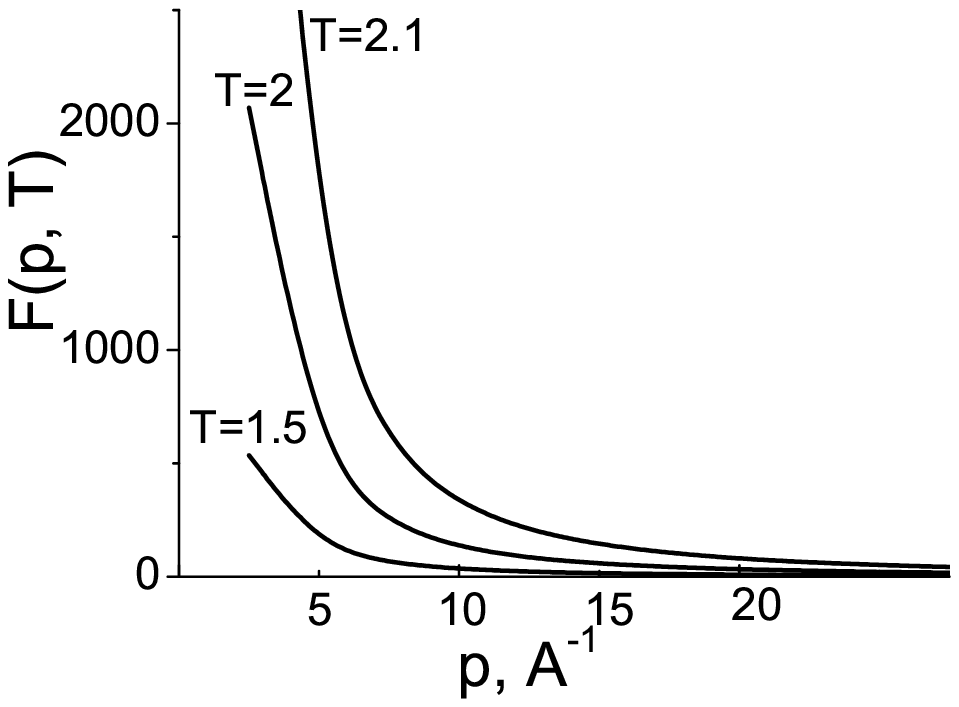}
\end{center}
\end{figure}

    Fig.~1.The momentum dependence of the temperature factor
(\ref{b19}), obtained for different temperatures.

\newpage

\begin{figure}
\begin{center}
\includegraphics{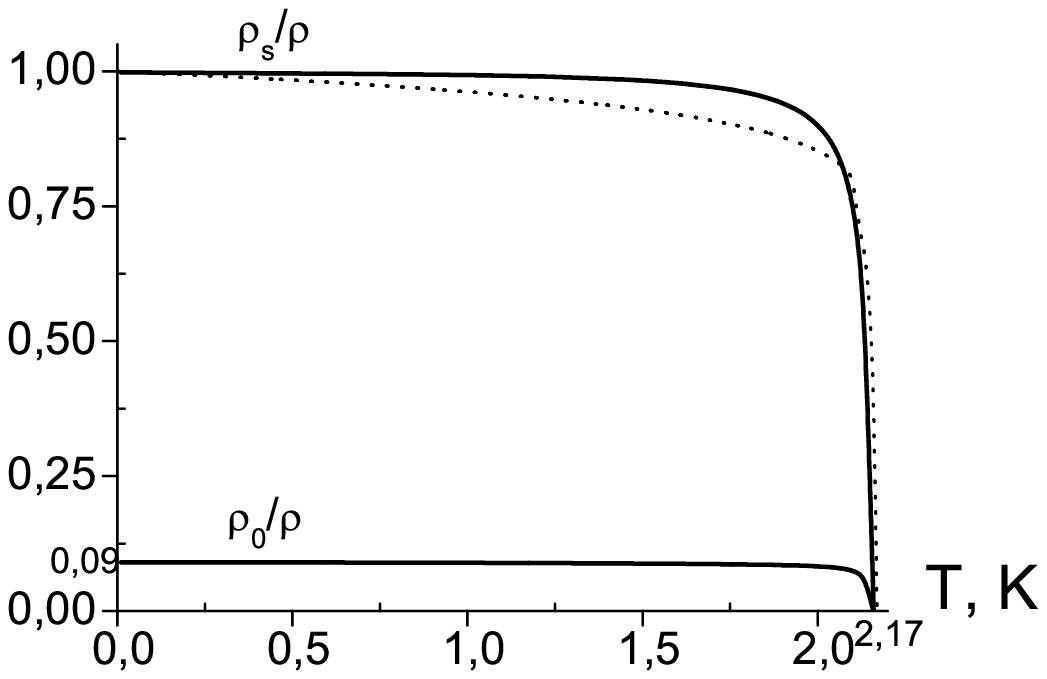}
\end{center}
\end{figure}

    Fig.~2. The temperature dependence of the density of BEC (lower
curve) and total density of superfluid component (upper curve),
obtained according  (\ref{b17}) and (\ref{b18}) for next
parameters $A_{11}=6.14 K,$ $D_{11}=2.03 K,$ $B_{11}=0.00018 K,$
$A_{12}=6.21 K,$ $D_{12}=3.12 K,$ $B_{12}=0.00225 K$. The
empirical temperature dependence of of superfluid component
\cite{20} (circles).

\end{document}